\documentclass[11pt,a4paper]{article}

\usepackage[affil-it]{authblk}
\usepackage[utf8]{inputenc}

\def\Lw{{\mathcal L}{}}
\def\tref{h_{(\rm r)}}

\def\Lbol{{\stackrel{\circ}{\mathcal L}}{}}

\def\ombol{{\stackrel{\circ}{\omega}}{}}
\begin{document}
\title{\textbf{Holographic Renormalization in Teleparallel Gravity}}
\author{Martin Kr{\v s}{\v s}{\' a}k\thanks{Electronic address: \texttt{krssak@ift.unesp.br}}}
\affil{\small{ Instituto de F\'isica Te\'orica, Universidade Estadual Paulista}\\\textit{
R. Dr. Bento Teobaldo Ferraz, 271, Barra Funda, 01140-070 -S\~ao Paulo SP, Brazil}}

\maketitle
\begin{abstract}
We consider the problem of IR divergences of the action in the  covariant formulation of  teleparallel gravity in  asymptotically Minkowski spacetimes. We show that divergences are caused  by  inertial effects and can be removed by adding an appropriate surface term, leading to the renormalized action. This process   can be viewed  as a teleparallel analogue of  holographic renormalization.  Moreover, we explore the variational problem in teleparallel gravity and explain how the variation with respect to the spin connection should be performed.
\end{abstract}

\section{Introduction} 
The variational principle plays a central role in theoretical physics. One of the necessary requirements for its consistency is the finiteness of the action. In general relativity, this is known to be a problem due to the presence of IR divergences, and is addressed using either the ``background subtraction" method \cite{Gibbons:1976ue} or holographic renormalization \cite{Balasubramanian:1999re,Skenderis:2002wp}. In the later one, IR divergences are cured by adding  appropriate surface terms to the action, which  in the original case of the asymptotically AdS spacetimes, could be interpreted as counterterms of a dual field theory through the AdS/CFT correspondence \cite{Maldacena:1997re}.
However, later this method turned out to work in other spacetimes  that are not known to possess  exact holographic duals \cite{Kraus:1999di, Balasubramanian:2001nb,Larsen:2004kf,Mann:2005yr}, suggesting that the  physical mechanism behind counterterms must be related to  some universal property of gravity, rather than to some specific realization of the holographic principle. An interesting work in  this direction is \cite{Miskovic:2009bm}, where the link with topological invariants was proposed.

In this paper, we  analyze the problem of IR divergences in the framework of teleparallel gravity, which is a well-known alternative formulation of general relativity  
that originates with  Einstein's unsuccessful attempt to find a unified field theory \cite{Einstein,Sauer}. Later \cite{Moller,Hayashi:1967se,Cho:1975dh,Hayashi:1977jd}, it was revived as a gravity theory only, without any unification purposes, that turned out to be equivalent to general relativity.   We call here this original approach---for a recent review, see \cite{Maluf:2013gaa}---as \textit{pure tetrad teleparallel gravity}, since the only variable is the tetrad, while the spin connection is assumed to vanish identically in all frames.  As a consequence, such  theory has a preferred frame and local Lorentz invariance is violated.

Recently,  it was discovered  that these problems can be avoided if  a nontrivial teleparallel spin connection satisfying the  vanishing curvature condition is considered. Use of such a connection leads to the covariant teleparallel gravity that respects local Lorentz symmetry. This covariant formulation of teleparallel gravity was first pioneered in \cite{Obukhov:2002tm} in the framework of metric-affine 
theories \cite{Hehl:1994ue}, and  later in \cite{Obukhov:2006sk,Lucas:2009nq}. For a textbook  overview, see \cite{AP}. 

The covariant formulation was fully developed 
in \cite{Krssak:2015rqa}, where it was shown that there exists a certain physically preferred way to choose a spin connection, which is naturally associated with the tetrad, and a method to extract this connection was proposed. The use of this spin connection removes the inertial effects  and a finite purely  gravitational action is obtained.  Therefore, finding the appropriate spin connection can be understood as a renormalization process, where  we remove  the spurious inertial effects from the action. 

In this paper, we  build upon the results of \cite{Krssak:2015rqa}, and show that the inertial effects represented by the spin connection enter the teleparallel action only through a surface term. Therefore, the action is renormalized by adding an appropriate surface term, what can be viewed  as an  analogue of holographic renormalization in teleparallel gravity.  Moreover, based on this construction, we show how to  perform the variation with respect to both the tetrad and the spin connection.

\section{The Covariant Formulation of Teleparallel Gravity}
Teleparallel gravity is formulated using  tetrad formalism, where the fundamental variable is the tetrad $h^a_{\ \mu}$\footnote{We use the convention, where  the Latin indices $a,b,...$ run over the tangent space, while the Greek indices $\mu,\nu,...$ run over spacetime coordinates. The spacetime indices are raised/lowered using the spacetime metric $g_{\mu\nu}$, while the tangent indices using the Minkowski metric $\eta_{ab}$ of the tangent space.}, related to the spacetime metric by
\begin{equation}
g_{\mu\nu}=\eta_{ab}h^a_{\ \mu} h^b_{\ \nu}.\label{met}
\end{equation}
In order to be able to  parallel transport tensors on manifold, we can define various spin connections; each characterized by its curvature and torsion.
In general relativity, we use  the torsionless Levi-Civita spin connection $\ombol^a_{\ b\mu}$,  which turns out to be completely determined from a tetrad, and the non-trivial geometry becomes characterized by the curvature tensor. The physical meaning of the  components of the Levi-Civita connection can be seen from  the geodesic equation
\begin{equation}
\frac{d u^a}{ds}+\ombol^a_{\ b\mu}u^b u^\mu=0,
\label{geo}
\end{equation}
where the spin connection plays a role of the total acceleration, which is due to both gravitational and inertial effects. The presence of the later can be easily  seen from the fact that the components of the spin connection may be non-vanishing in absence of gravity and change under local Lorentz transformations. The curvature tensor is  constructed from derivatives of the connection and plays the physical role of the tidal tensor that is of  purely gravitational origin.

In  teleparallel gravity, we use a different connection defined by the condition of vanishing curvature
\begin{equation}
R^{a}_{\,\,\, b\mu\nu}(\omega^a_{\ 
b\mu})= \partial_\mu\omega^{a}_{\,\,\,b\nu}-
\partial_\nu\omega^{a}_{\,\,\,b\mu}
+\omega^{a}_{\,\,\,c\mu}\omega^{c}_{\,\,\,b\nu}-\omega^{a}_{\,\,\,c\nu}
\omega^{c}_{\,\,\,b\mu}\equiv 0,
\label{curvv}
\end{equation}
 which is satisfied by a pure gauge-like  connection \cite{AP}
\begin{equation}
\omega^a_{\ b\mu}=\Lambda^a_{\ c} \partial_\mu \Lambda_b^{\ c},\label{connection}
\end{equation}
which is completely determined by the Lorentz matrix, and hence is related to inertial effects only. It is sometimes referred to as the \textit{purely inertial connection} \cite{AP}.

The teleparallel connection (\ref{connection}) has non-vanishing torsion
\begin{equation}
T^a_{\ \mu\nu}(h^a_{\ \mu},\omega^a_{\ b\mu})=
\partial_\mu h^a_{\ \nu} -\partial_\nu h^a_{\ \mu}+\omega^a_{\ b\mu}h^b_{\ \nu}
-\omega^a_{\ b\nu}h^b_{\ \mu},
\label{tordef}
\end{equation}
and can be related to the Levi-Civita connection using  Ricci theorem
\cite{Nakahara} 
\begin{equation}
\omega^a_{\ b \mu}=\ombol^a_{\ b \mu} + K^a_{\ b \mu},
\label{decomp}
\end{equation}
where 
\begin{equation}
K^{a}_{\  b\mu}=\frac{1}{2}
\left(
T^{\ a}_{\mu \ b}
+T^{\ a}_{b \ \mu}
-T^{a}_{\  b\mu}
\right),
\label{contortion}
\end{equation}
is the contortion tensor. 

The Lagrangian density of teleparallel gravity is given by  
\begin{equation}
\Lw=\frac{h}{4 \kappa} T^a_{\ \mu\nu}S_a^{\ \mu\nu}, \label{lagtot}
\end{equation}
where $h=\det h^a_{\ \mu}$, $\kappa=8\pi G$ is the gravitational constant (in $c=1$ units), and $S_a^{\ \mu\nu}$ is the superpotential defined by
\begin{equation}
S_a^{\ \mu\nu}=K^{\mu\nu}_{\ \ a}
-h_a^{\ \nu}T^{ \mu}
+h_a^{\ \mu}T^{ \nu},
\label{sup}
\end{equation}
where $T^\mu=T^{\nu\mu}_{\ \ \ \nu}$ is the vector torsion.
The  vacuum field equations are obtained by varying (\ref{lagtot}) with respect to the tetrad, and is given by
\begin{equation}
h^{-1}\partial_\sigma \left(h S_a^{\ \rho\sigma} \right)-
h_a^{\ \mu} S_b^{\ \nu\rho} T^b_{\ \nu\mu}+
\frac{h_a^{\ \rho}}{4} T^b_{\ \mu\nu}S_b^{\ \mu\nu}
 -  \omega^b_{\ a\sigma}S_b^{\ \rho\sigma}
= 0.\label{femix}
\end{equation}
The variation with respect to the spin connection vanishes identically. We postpone a proof of this statement to the following section, where the central result of this paper, represented by eq. (\ref{rel}), is   derived.

The equivalence of teleparallel gravity and general relativity follows from a  straightforward calculation using Ricci theorem (\ref{decomp}) that shows that the Lagrangian (\ref{lagtot}) is equivalent (up to a surface term) to the Einstein-Hilbert Lagrangian
\begin{equation}
\Lw =\Lbol_{EH}-\partial_\mu \left(\frac{h}{\kappa} \, T^\mu \right), \label{lagequiv}
\end{equation}
implying the equivalence of the field equations of teleparallel gravity (\ref{femix}) and  general relativity.

To understand the  physical meaning of the teleparallel spin connection, let us recall Ricci theorem (\ref{decomp}) and interpret it as a decomposition of the Levi-Civita connection into the teleparallel spin connection and the contortion tensor. Obviously, this decomposition is not unique since there are many possible ways to choose the observer (represented by some $\Lambda^a_{\ b}$), each corresponding to a different spin connection of the form (\ref{connection}).  For example, in the pure tetrad teleparallel gravity \cite{Maluf:2013gaa}, where the teleparallel spin connection vanishes identically in all frames, the torsion tensor 
\begin{equation}
T^a_{\ \mu \nu} (h^a_{\ \mu},0)
\label{torun}
\end{equation}
plays the role of the  total acceleration and  represents both gravity and inertia.

However, in the covariant formulation of teleparallel gravity with the non-trivial connection (\ref{connection}), it  turns out to be possible to split the Levi-Civita connection in a physically desirable way, where  the teleparallel  connection  represents the inertial effects and the contortion tensor  represents gravity only.
The  self-consistent procedure to perform  this split was proposed in \cite{Krssak:2015rqa}, which runs according to the following scheme.
We start with a tetrad $h^a_{\ \mu}$ obtained as a solution to the field equation with the zero spin connection, and  define  the \textit{reference tetrad} $\tref{}^a_{\ \mu}$ by setting some parameter that controls the strength of gravity to zero. In practice, we can either choose the gravitational constant to vanish
\begin{equation}
\tref^{\;a}{}_{\mu}\equiv\left. h^a_{\ \mu}\right|_{G\rightarrow 0}. \label{reftet}
\end{equation}
or consider the asymptotic limit $r\rightarrow \infty$, where gravity is expected to vanish in the case of the asymptotically Minkowski spacetimes.

Since we want torsion to represent gravity only, we  require
torsion to vanish in its absence  
 \begin{equation}
 T^a_{\ \mu\nu}(\tref^{\;a}{}_{\mu},\omega^a_{\ b\mu})\equiv 0, \label{cond}
\end{equation}
what completely determines the spin connection in terms of the reference tetrad (\ref{reftet}) as 
\begin{equation}
\omega^a_{\ b\mu} = \ombol^a_{\ b\mu}(\tref). \label{leveq}
\end{equation}
Then, the torsion tensor constructed out of the ``full" tetrad and the spin connection
\begin{equation}
T^a_{\ \mu \nu} (h^a_{\ \mu},\omega^a_{\ b\mu}) \label{gtor}
\end{equation}
represents purely gravitational torsion with the spurious inertial contributions removed \cite{Krssak:2015rqa}.

To elaborate on the choice of  the reference tetrad we point out that the reference tetrad is related to the Minkowski spacetime through the relation (\ref{met}). Therefore, various reference tetrads represent different physical observers in the Minkowski spacetime. There exists  a  simple diagonal tetrad in the Cartesian coordinate system for which  the connection (\ref{leveq}) vanishes identically. All the other reference tetrads are related  to it by a coordinate change and/or a local Lorentz transformation. If they are related by a local Lorentz transformation, a non-trivial spin connection is generated on the account of non-tensorial transformation properties of the spin connection under local Lorentz transformations. This insight can be used to determine the proper reference tetrad in the case when the limit procedure (\ref{reftet}) is not easily achievable.


%
%
%
%
\section{Surface Terms in the Teleparallel Action\label{surf}}
 We expect the action  constructed from the torsion tensor with vanishing spin connection (\ref{torun}) to be IR divergent due to the presence of the inertial effects. The inertial forces are just fictitious forces and, unlike the ``real" forces associated with physical fields,   do not necessarily vanish at infinity \cite{Landau:1982dva}. Since the action  $S(h^a_{\ \mu},0)$ is an integral of such spurious contributions  over the whole spacetime, it is natural to expect it to be IR divergent.
 
On the other hand, if the action is constructed from the purely gravitational  
torsion (\ref{gtor}), with the spurious inertial contributions removed using the appropriate spin connection (\ref{leveq}), we obtain the purely gravitational action $S(h^a_{\ \mu},\omega^a_{\ b\mu})$  that we expect to be IR finite.

The primary task of this paper is to find an explicit relation between these actions\footnote{We should recall here that there exists a class of frames called \textit{proper frames}, in which the  spin   connection vanishes, and these actions coincide \cite{Krssak:2015rqa}. See \cite{Krssak:2015oua} for an explicit example. Our task here is to derive the relation between actions in the case of the non-vanishing spin connection.}. In principle, it should be possible to achieve this by a straightforward calculation: writing both actions in terms of tetrad and spin connection explicitly,  isolating the contributions of the spin connection, and using the zero curvature condition (\ref{curvv}), which introduces teleparallelism and relates terms squared in connection to terms linear in their derivatives.

However, it turns out to be possible to achieve the same result in a simpler way,  using the equivalence of the teleparallel and the Einstein-Hilbert actions (\ref{lagequiv}). Since the Levi-Civita connection of general relativity  is a function of a tetrad, the Einstein-Hilbert action can be expressed in terms of a tetrad only. Therefore, the same Einstein-Hilbert action corresponds to both $S(h^a_{\ \mu},0)$ and $S(h^a_{\ \mu},\omega^a_{\ b\mu})$, and we can write
\begin{equation}
\Lw(h^a_{\mu},\omega^a_{\ b\mu}) + \partial_\mu \left[\frac{h}{\kappa} \, T^\mu(h^a_{\mu},\omega^a_{\ b\mu})\right] =
\Lw(h^a_{\mu},0) + \partial_\mu \left[\frac{h}{\kappa} \, T^\mu(h^a_{\mu},0) \right].
 \end{equation}
Taking the definition of the torsion tensor (\ref{tordef}), and contracting it with $h_a^{\ \nu}$, we  find 
\begin{equation}
T_{\mu}(h^a_{\ \mu},\omega^a_{\ b\mu})=h_a^{\ \nu}
\partial_\mu h^a_{\ \nu} -h_a^{\ \nu} \partial_\nu h^a_{\ \mu}
-h_a^{\ \nu} \omega^a_{\ b\nu}h^b_{\ \mu}=
T_{\mu}(h^a_{\ \mu},0)
-\omega_\mu,
\end{equation}
where we have defined $\omega_\mu=\omega^{a}_{\ b \nu}h_a^{\ \nu}h^b_{\ \mu}$. Combining these two relations, we find that the Lagrangians are related by a total divergence only
\begin{equation}	
\Lw (h^a_{\ \mu},\omega^a_{\ b\mu})=
\Lw (h^a_{\ \mu},0) + \frac{1}{\kappa} \partial_\mu \left(
h \omega^\mu
\right). \label{rel}
\end{equation}
This ``holographic" relation is the central result of this paper.  
It shows that  the divergences are removed from the action by adding an appropriate surface term to the action, analogously to holographic renormalization. However, in teleparallel gravity it can be interpreted as a removal of the spurious inertial effects from the theory.

This relation allows us to address the variational problem in the covariant teleparallel gravity properly. Since the Lagrangian (\ref{lagtot}) is a function of both the tetrad and the  spin connection,  the variation with respect to both variables should be considered.  The variation with respect to the tetrad leads straightforwardly to the field equations (\ref{femix}), but the variation with respect to the  spin connection is a  more tricky task. To illustrate this problem, let us naively consider  a straightforward variation of the Lagrangian (\ref{lagtot}) with respect to the spin connection. We can observe  that the spin connection appears in the action without derivatives, and it is easy to find out that the variation  leads to $S_a ^{\ \mu\nu}=0$, implying a trivial theory.
The reason for this puzzling result is that  we are naively varying a Lagrangian quadratic in a general torsion,  not the teleparallel one. 

In order to obtain the correct result, we should keep in mind that the theory with the  Lagrangian (\ref{lagtot}) is a teleparallel theory if and only if we  use the teleparallel condition (\ref{curvv}),  solved by the connection (\ref{connection}). The teleparallel condition (\ref{curvv}) relates the terms quadratic in the connection  to their derivatives, which turn out to form a total derivative, leading to the relation (\ref{rel}). Since the spin connection enters the action only through the surface term, the  variation with respect to the spin connection vanishes 
\begin{equation}
\frac{\delta \Lw (h^a_{\ \mu},\omega^a_{\ b\mu})}{\delta \omega^a_{\ b\mu}}=0.
\end{equation}
Moreover, it implies that the field equations (\ref{femix}) derived from $ \Lw (h^a_{\ \mu},\omega^a_{\ b\mu})$ and  $\Lw (h^a_{\ \mu},0)$ are the same, what  allows us to solve the field equations with the vanishing spin connection. This is crucial for the procedure of determining the spin connection proposed in \cite{Krssak:2015rqa}, since the solution is needed to define the reference tetrad (\ref{reftet}), and consequently the correct spin connection (\ref{leveq}). Furthermore, this result explains why in pure tetrad  teleparallel gravity the solutions to the field equations are correct \cite{Moller,Hayashi:1967se,Cho:1975dh,Hayashi:1977jd,Maluf:2013gaa}. It  illustrates  actually the irrelevance of the teleparallel spin connection for obtaining the solution of the field equations.

\subsection*{Schwarzschild Solution}
To  illustrate the renormalization process, let us consider the diagonal tetrad representing  the Schwarzschild solution in the spherical coordinate system
\begin{equation}
h^a_{\ \mu}=\textrm{diag}\left( \sqrt{f(r)},
1/\sqrt{f(r)},r, r\sin \theta \right),
\label{tetdiag}
\end{equation} 
where
\begin{equation}
f(r)=1-2GM/r. \label{alphasol}
\end{equation}
The unrenormalized Lagrangian density  is given by 
\begin{equation}
\Lw(h^a_{\ \mu},0)=\frac{1}{\kappa}\sin\theta. \label{schwactnon}
\end{equation}
The non-vanishing components of the spin connection corresponding to the diagonal tetrad (\ref{tetdiag}) are 
\begin{equation}
\omega^{\hat{1}}_{\ \hat{2}\theta}=-1, \quad 
\omega^{\hat{1}}_{\ \hat{3}\phi}=-\sin\theta, \quad
\omega^{\hat{2}}_{\ \hat{3}\phi}=-\cos\theta. \label{spschw}
\end{equation}
We find the contraction
\begin{equation}
\omega^\mu=\left(0, -\frac{2f}{r},-\frac{\cot\theta}{r^2},0\right), \label{wu}
\end{equation}
and construct the surface term
\begin{equation}
\partial_\mu (h \omega^\mu)=
\left[1+2 \frac{(GM-r)}{fr}\right]\sin\theta.
\end{equation}
We can check that adding  this surface term to  the unrenormalized action (\ref{schwactnon}) according to the relation (\ref{rel}) leads to the renormalized Lagrangian
\begin{equation}
\Lw(h^a{}_{\mu},\omega^a_{\ b\mu})=
\frac{2}{\kappa}
 \left[1+\frac{(GM-r)}{r\sqrt{f}} \right]
\sin\theta,
\end{equation}
and the renormalized action \cite{Krssak:2015rqa}.
\section{ Conclusions}
In this paper we have analysed the problem of IR divergences of the action in the covariant formulation of teleparallel gravity in asymptotically Minkowski spacetimes. We have shown that, since torsion  is related to the acceleration of the observer, there is a possible mixing of  gravitational and inertial effects.  The inertial effects are not related to the actual physical fields, and hence do not necessarily vanish at infinity, what leads to IR divergences of the teleparallel action. This is an important physical insight provided by teleparallel gravity   that the IR divergences of the action are actually physically meaningful and are related to  inertial effects.

These spurious inertial contributions can be removed, provide the appropriately chosen teleparallel spin connection (representing the inertial effects in the tetrad) is used. We have then analysed how this spin connection enters the teleparallel action, what led us to the "holographic" relation (\ref{rel}) that shows that  the spin connection enter the action only through the surface term. Therefore, the removal of the spurious inertial contributions can be understood as an addition of the appropriate surface term that renders the action finite, what  can be considered as a teleparallel  analogue of holographic renormalization.

An interesting aspect of  renormalization  of the teleparallel action is that it has features of both holographic renormalization and background subtraction methods. The analogy with background subtraction  was already suggested in \cite{Krssak:2015rqa}, where it was observed that the spin connection is calculated from the reference tetrad, representing the reference or  background spacetime. The current  work provides further insight that, for the purpose of renormalization of the action, the reference tetrad does not have to be defined everywhere. Since  it enters the action only through the surface term,  it is sufficient if the reference  tetrad is properly defined  at infinity.

The  relation (\ref{rel}) not only explains how the action is renormalized, but  is of crucial importance for the consistency of the variational principle in  the covariant formulation of teleparallel gravity. Since the spin connection is  a function of the reference tetrad only, it should be considered as an independent variable. Therefore, the variation of the action with respect to the spin connection should be considered as well. The relation (\ref{rel}) allows us to perform this variation easily, to find out  that this variation vanishes and the field equations are independent of the choice of the spin connection. This result explains the success of the previous, non-covariant, pure tetrad  formulations in finding solutions to teleparallel gravity, and clarify the relation with the  covariant formulation considered in this paper.

\section{Acknowledgments}
The author would like to thank Jos\'e Geraldo Pereira for a fruitful discussion and comments on this manuscript. This work was financially supported by  FAPESP.

\end{document}